\documentclass{article}
\usepackage[utf8]{inputenc}
\usepackage{amsmath}
\usepackage{amssymb}
\usepackage{amsthm}
\usepackage{amscd}
\usepackage{amsfonts}
\usepackage{setspace}
\usepackage{graphicx}%
\usepackage{color, url, wrapfig, hyperref, xcolor, soul}
\usepackage{titlesec}
\usepackage{mdframed} 

\usepackage[font=small,labelfont=bf]{caption}
\usepackage[left=0.75in,right=0.75in,top=0.75in,bottom=0.75in]{geometry}
\usepackage{enumitem}

\usepackage[sort,round]{natbib}
\usepackage{times}

\titlespacing{\section}{0pt}{1.1ex}{1.1ex}
\titlespacing{\subsection}{0pt}{1ex}{0.7ex}

\title{Employing Social Media to Improve Mental Health Outcomes}
\author{Munmun De Choudhury}
\date{
        School of Interactive Computing\\
        Georgia Tech\\
        Atlanta, GA 30332\\
        munmund@gatech.edu}

\begin{document}

\begin{mdframed}[linecolor=black, linewidth=0.5pt, roundcorner=5pt, innermargin=10pt, outermargin=10pt]
\small
\begin{spacing}{1.2}
\textbf{This is the preprint of a chapter that will appear in the Handbook of Computational Social Science edited by Taha Yasseri, forthcoming 2025, Edward Elgar Publishing Ltd.}

\vspace{0.3cm}
\noindent \textbf{Suggested citation:} De Choudhury, Munmun (2025). ``Employing Social Media to Improve Mental Health Outcomes." In: T. Yasseri (Ed.), \textit{Handbook of Computational Social Science}. Edward Elgar Publishing Ltd.
\end{spacing}
\end{mdframed}

\vspace{0.1cm}


\begin{center}
    \LARGE \textbf{Employing Social Media to Improve Mental Health Outcomes} \\[10pt]
    \large Munmun De Choudhury \\[5pt]
    \normalsize School of Interactive Computing \\ Georgia Tech \\ Atlanta, GA 30332 \\ munmund@gatech.edu
\end{center}

\vspace{0.5cm}

\paragraph{Abstract} 
As social media platforms are increasingly adopted, the data the data people leave behind is shining new light into our understanding of phenomena, ranging from socio-economic-political events to the spread of infectious diseases. This chapter presents research conducted in the past decade that has harnessed social media data in the service of mental health and well-being. The discussion is organized along three thrusts: a first that highlights how social media data has been utilized to detect and predict risk to varied mental health concerns; a second thrust that focuses on translation paradigms that can enable to use of such social media based algorithms in the real-world; and the final thrust that brings to the fore the ethical considerations and challenges that engender the conduct of this research as well as its translation. The chapter concludes by noting open questions and problems in this emergent area, emphasizing the need for deeper interdisciplinary collaborations and participatory research design, incorporating and centering on human agency, and attention to societal inequities and harms that may result from or be exacerbated in this line of computational social science research.

\paragraph{Keywords:} ethics; machine learning; mental health; prediction; social media 

\section{Introduction}

Mental and psychological well-being concerns like mood disorders can have wide-reaching effects on people's daily activities, education, employment, occupational functioning, and relationships~\citep{kessler2003epidemiology}. One in four adults, or 61.5 million Americans, are reported to experience such a challenge in a given year~\citep{mathers2006projections,reeves2011mental}. Besides being directly debilitating to sufferers, mental illness can adversely affect chronic health conditions, such as cardiovascular disease, cancer, diabetes, and obesity~\citep{robson2007serious}. 
Although a number of care programs have been devised for its detection and treatment, the majority of the millions of Americans who meet mental illness criteria are untreated or under-treated~\citep{ten2013lifetime}. Furthermore, ethnically minoritized groups such as Blacks and Hispanics are significantly less likely to receive therapies than are other ethnic groups~\citep{primm2010role,cleary2014marginalization}. Importantly, individuals struggling with mental illness tend to be ones who are disproportionately marginalized in the society, facing hurdles right from the early phases of child development to being valuable participants of the society~\citep{scrutton2017epistemic}.

Naturally, the question of how to address mental health issues has existed since antiquity; just that the answers have evolved across cultures and millennia, adapting as the understanding of the human condition has changed in the face of advances in science, chemistry, medicine, and psychology~\citep{hardy2019beautiful}. If we are to look back, humans have tried since the past 7,000 years to cure and treat the mentally ill~\citep{foerschner2010history}. Psychoanalysis, the first scientific paradigm, originated at end of 19th century in renowned psychologist Sigmund Freud's work called ``Project for a Scientific Psychology,'' the goal of which was to ``furnish a psychology that will be a natural science''~\citep{freud1977introductory}. With the invention of antidepressants and eventually anti-psychotic drugs in the mid and late 20th century, two paradigms emerged: Clinicians and researchers began to use a ``biological'' psychiatry paradigm which described mental disorders as dysfunction of the brain physiology~\citep{trimble2010biological}, while learning behavioral and cognitive scientists began to regard the learner as an active interpreter of the situation, giving rise to the widely acclaimed Cognitive Behavioral Therapy by Aaron Beck -- an approach used even today for changing patterns of thought that causes disturbed emotional behaviors~\citep{butler2006empirical}. 

Indeed, with these paradigms, we have come a long way from relieving people of ``evil spirits'', putting people in asylums, or conducting lobotomies~\citep{dvoskin2020brief} -- however, given a lack of unification between these two paradigms subject to the mind-body dualism~\citep{raese2015pernicious}, alongside given widely prevalent social stigma, systemic barriers, and deep-seated disparities~\citep{corrigan1998impact}, success in efforts to help individuals with mental illness continues to evade us. In this light, in the language of Thomas Kuhn~\citep{kuhn1970structure} -- could advances in the Computer Science field support the next ``paradigm shift'' in mental health? In this chapter, drawing on my research over the past decade, we will explore attempts to answer this question.

Many mental health challenges are known to be characterized by latent processes that include negative perspectives, self-focused attention, loss of self-worth and self- esteem, and social disengagement~\citep{thoits1999sociological,de2005social}. Today, with social media, many of these latent processes, such as the context and content of such as one’s affective, behavioral, and cognitive reactions are recorded in the present and can be observed longitudinally and unobtrusively~\citep{golder2011diurnal}. As many of these platforms increasingly gain traction among individuals~\citep{auxier2021social}, they are building up large user bases, including many people who have been using the service for years. At the same time, individuals are increasingly appropriating social media platforms to engage in candid disclosures of various well-being challenges~\citep{de2014mental,lin2016health}. 
Together, these provide evidence of how social media platforms might be emergent tools to understand and learn about people’s mental and behavioral health, so as to provide them with appropriate resources to cope with this stigmatizing condition, as well as with timely and tailored interventions to manage this psychological challenge. 

The research presented in this chapter builds on this opportunity provided by emergent trends in the computational social science field where social media data has been appropriated to understand several nuanced aspects of human behavior, whether the diffusion of information, the evolution of communities, crisis mitigation, and sociopolitical phenomena~\citep{lazer2009computational,lazer2020computational}. The work discussed here is organized in three thrusts: 

\begin{description} 
    \item[Thrust I] If we can employ social media to proactively detect one’s risk to mental health challenges 
    \item[Thrust II] What are appropriate models of collaboration to enable translation of these detection algorithms in the real-world 
    \item[Thrust III] What types of ethical considerations and 
    privacy-preserving mechanisms need to underlie this research to ensure that the benefits of the work, in view of the former two thrusts, outweigh the risks posed to relevant stakeholders. 
\end{description}

The primary motivation behind this research agenda is grounded in the Social Ecological Model (SEM)~\citep{wold2018health},  which emphasizes (mental) health to be deeply embedded in the complex interplay between an individual, their relationships, the communities they belong to, and societal factors. The emotion and language used by individuals in their social media postings may indicate feelings of worthlessness, guilt, helplessness, and self-hatred that characterize many mental illnesses as manifested in everyday life~\citep{berkman2000social}. Additionally, sufferers of mental illness often show withdrawal from social situations and activities -- in other words, the etiology of these conditions typically includes social environmental factors~\citep{compton2015social}. Characterization of social media activity and
changing social ties within social media can provide measurement of such withdrawal and capture the sufferers’ social context in a manner that might provide deep insights into their experiences and condition~\citep{de2012not}. The SEM framework thus provides a guiding conceptualization to employ social media data in the service of improving mental health outcomes.

\section{Predicting Mental Health States with Social Media}

Self-reports have formed the primary basis and hallmark in the detection and diagnosis of mental illnesses~\citep{hariman2019future}. However the large temporal gaps across which these measurements are made, as well as the limited number of participant responses makes it difficult to track and identify risk factors that may be associated with mental illness, or to develop effective intervention programs. In contrast to surveys and questionnaires, where responses are prompted by the experimenter and typically comprise recollection of (sometimes subjective) health facts, social media measurement of behavior captures social activity and language expression in a naturalistic setting. Such activity is real-time, and happens in the course of a person’s day-to-day life~\citep{insel2018digital}. 
Harnessing this opportunity, a large body of research has sought to employ retrospective social media data of individuals to build machine learning predictive models that can detect at-risk individuals or at-risk mental health episodes (e.g., perspective piece by ~\cite{inkster2016decade}).

Some of the earliest work in this area provided valuable insights into how social media enables a passive way and also serves as a  ``verbal'' sensor to gather quantifiable signals about the social-ecological dimensions relating to an individual's mental health~\citep{de2014can}. For example, in a first of its kind study~\citep{de2013major,de2013postpartum,dechoudhury2014fbppd} we devised methods to identify linguistic and emotional correlates for postpartum depression among new mothers who used Twitter and Facebook.  We found that behavioral concerns such as postpartum depression may be reflected in new mothers’ social media use: including lowered positive affect and raised negativity, and use of greater first person pronouns indicating higher self-attentional focus. In fact, the behavioral changes of these new mothers could be predicted by leveraging their activity from simply the prenatal
period. Taking on this unique opportunity provided by social media, thereafter, we developed predictive models that could identify whether an individual could be at risk of depression, prior to the onset of the condition. Specifically, we utilized a panel of nearly 400 individuals and their entire archives of Twitter data to achieve a predictive accuracy of over 71\%, significantly improving over competing baselines~\citep{de2013predicting}. 
Our models could assess this risk not only prior to the self-reported onset of depression, but also compared favorably to 
individual-level assessments of depression made by validated screening instruments, as well as population-level prevalence statistics
~\citep{de2013social}.

Subsequent research bolsters and further validated these initial investigations. a) We used Reddit data and transfer learning to automatically assess levels of acute stress following campus incidents of gun violence~\citep{saha2017modeling}; b) We employed social media (Twitter) data as a way to computationally detect which individuals are likely to have schizophrenia, by combining the use of machine learning techniques and clinical appraisals ~\citep{birnbaum2017collaborative}; c) We leveraged Facebook and ecological momentary assessments in a semi-supervised learning framework to predict mood instability in individuals with bipolar or borderline personality disorders ~\citep{saha2017CL};  d) We harnessed self-disclosures made on pseudonymous Reddit communities to identify mental health risk among LGBTQ+ identifying individuals, drawing upon the minority stress theory~\citep{saha2019language}; and e) We assessed levels of mental health and psychosocial concerns during the COVID-19 pandemic, based on a quasi-experimental study of geocoded Twitter data~\citep{saha2020psychosocial}.  

The above formative research showcased how records of people's activity on social media can enable machine learning algorithms that can assess and even predict risk to a variety of mental health challenges~\citep{de2018integrating,coppersmith2018natural}. However, while opening up new opportunities of further investigation, this research also raised many important questions: \textit{a) what is the clinical, real-world, and theoretical validity of the developed models and how do we guarantee that; b) how does social media data and the inferences derived from it supplement and complement offline behavioral and health data; and c) whether social media activity and language could signal causal relationships with mental health.} 
In the subsequent subsections, I describe research we conducted in response to these critical directions. 

\subsection{Validity of Social Media-Powered Algorithms}

Targeting the first question, we discovered gaps in clinical utility and psychometric (construct) validity of social media predictions of mental health, due to an over-reliance on behavioral proxies instead of clinically assessed ground truth and over-zealous attention to scaling machine learning models at all costs~\citep{ernala2019methodological}. 

We started by looking at the myriad machine learning algorithms developed in prior research that have attempted to predict mental health states of individuals from their social media feeds. An implication of this research lies in informing evidence-based diagnosis and treatment~\citep{torous2014promise}. The success of these algorithms, therefore, hinges on access to ample and high-quality gold standard labels for model training~\citep{mohr2017personal}. In mental health, gold standard labels often comprise diagnostic signals of people’s clinical mental health states, for instance, whether an individual might be suffering from a specific mental illness, or at the cusp of experiencing an adverse episode like a relapse or suicidal thoughts~\citep{birnbaum2015role}.

However, unlike conventional machine learning tasks in fields like computer vision and natural language processing, extensive gold standard data for predicting clinical diagnoses of mental illnesses from social media is not readily available. Literature has advocated the use of clinically validated diagnostic information collected from patient populations for building such predictive models~\citep{rutledge2019machine}. However, undertaking such efforts presents many practical and logistical challenges. These range from the difficulties in recruiting a sensitive and high risk population, to the myriad privacy and ethical concerns that accompany engaging directly with vulnerable individuals. Because of the effort- and time-consuming nature of such data acquisition approaches and the need for deep-seated cross-disciplinary partnerships, particularly with clinicians, researchers have noted such data acquisition efforts to not scale easily and quickly to large populations~\citep{coppersmith2014quantifying}.

Consequently, researchers have operationalized online behaviors as diagnostic signals to build machine learning approaches that predict mental illness diagnoses. These ``proxies'' are easily accessible and inexpensively gathered, without the need to directly engage with the individuals themselves. One notable example from
literature consists of public self-reports of mental illnesses made by individuals in their social media feeds~\citep{coppersmith2018natural}. Adopting a model triangulation approach and comparing three proxies with clinically-gauged mental health status of a set of patients through building binary classification models, we found that although the models trained on proxy diagnostic signals demonstrated strong internal validity, they performed poorly on the clinically diagnosed patient data, thus suffering from poor external validity. A deep dive in the performance of these classifiers via an error analysis revealed several methodological gaps in the way these proxy diagnostic signals are conceived and employed in the predictive frameworks. These gaps range from uncertainties in the construct validity of the proxy signals, and poor theoretical grounding, to a variety of population and data sampling biases.

This research concluded by providing guidelines for researchers engaging with prediction of mental health states from social media data, and for clinicians interested in incorporating such machine learning based diagnostic assessments in clinical decision-making. These guidelines include the use of deeper computing and clinical research partnerships and investments in building benchmarked social media datasets, annotated with clinical information, through transparent and voluntary consent of individuals donating their data.

\subsection{Supplementing and Complementing Conventional Signals}

In answering the second question posed above, we  integrated and harmonized diverse data streams, beyond social media alone, to quantify mental health outcomes~\citep{mattingly2019tesserae}. Our motivation stems from the fact that while social media can be collected in an inexpensive and unobtrusive way at scale, it, by itself, is highly homogeneous and does not reveal a comprehensive picture of an individual’s social ecological state. It has been argued that augmenting it with other forms of social-behavioral data can help overcome the biases and limitations of social media data alone. With this observation and challenge in mind, we proposed a machine learning framework~\citep{saha2019imputing} that used noisy and incomplete data derived from social media, wearable sensors, smartphones, and bluetooth location trackers, over an year spanning 750 individuals who contributed over 300K Facebook posts to assess physical and physiological behavior, as well as psychological states and traits. This integration framework used state of the art imputation approaches, multimodal data fusion, ensemble learning, and representational learning to successfully predict self-reported mood, stress, and anxiety ($\rho=$0.3-0.4). 

While this research allowed us to establish how social media data can work in concert with other forms of sensed data, it did not provide insights into the gains it offered over conventional sources. To address this, in subsequent research we developed new methodologies and analytical frameworks that showed that both person-centered analyses of social media data and passive sensor data-augmented social media predictions provided gains over variable-centered predictions that used social media alone~\citep{saha2021person,robles2020jointly}. 
Since precision medicine argues that such predictions be personalized to account for clinical heterogeneity, we thereafter augmented the predictive assessments by first identifying clusters of individuals based on physical sensor data streams, and then building personalized predictive models on these stratified samples, where each stratified cluster represented a different set of lifestyles and behaviors~\citep{das2019multisensor,saha2021person}. We found these person-centered models to improve over variable-centered predictions by 4-78\%.
This made the case not only for adopting personalized machine learning in deriving mental health assessments, but also that ecological factors, that could gained from complementary data sources, need to be accounted for in these predictive approaches.

\subsection{Understanding Causality}

Seeking to answer the question above around causality, we identified linguistic shifts in online discourse causally associated with emerging suicidal ideation (SI)~\citep{de2016discovering}. Focusing on  mental health support communities on Reddit, we applied propensity score matching to explore how users may express suicidal ideation in the future, while controlling for the historical use of linguistic constructs of mental health. Through statistical analysis methods developed for causal inference, we isolated the effects of linguistic constructs from observed confounding factors, which allowed us to examine differences between those who proceed to discuss suicidal ideation in the future, from those who do not. We discovered transition to SI to be associated with psychological attributes like heightened self-attentional focus, poor linguistic coherence, reduced social engagement, and manifestation of hopelessness, anxiety, impulsiveness and loneliness. Finally, we examined whether we can automatically predict the tendency of individuals discussing mental health concerns to engage in these characteristic behaviors -- a logistic regression classifier yielded an accuracy of 81\%, against a 50\% chance in predicting future SI.

In a closely related work, we showed, also via the use of a propensity score based matching framework, how social media data may include protective factors against suicidal risk~\citep{de2017language}. Online social support is known to play a significant role in psychological well-being~\citep{eysenbach2004health,schwarzer2007functional}, however this link has rarely been well quantified, due to the paucity of longitudinal, pre- and post risk exacerbation data, and reliable methods that can examine causality between past availability of support and future risk. Therefore, we proposed a computational method to measure how the language of comments in Reddit mental health communities influences risk to suicidal ideation in the future, as demonstrated in the posting activities of individuals in these communities. 
Incorporating human assessments in a stratified propensity score analysis based framework, we identified comparable subpopulations of individuals in Reddit mental health communities (like above) and measured the effect of online social support language. On interpreting these linguistic cues with Suhr et al's~\citep{suhr2004social} model of social support (``Social Support Behavioral Code''), we found that esteem and network support play a prominent protective role in reducing forthcoming risk of expressions of SI. 

Summarily, we were successful in discovering specific linguistic cues as well as supportive conversations on social media that increased or decreased forthcoming risk of suicidal thoughts.
 
\section{Participatory Approaches to Enable Real-World Use} 

Given the plethora of opportunities put forth by the above described research, when can these developed predictive algorithms be considered ready for real world use? To answer this question, it is important to note that most of these algorithms have been developed opaquely and isolatedly, without involvement of stakeholders who are the most affected by them. In the absence of adequate stakeholder involvement and feedback, such algorithms, when deployed in the real world, can lead to mistrust and poor perceptions of accountability, ill-informed folk theories behind how they work, violated expectations, and reinforced biases. That is, the extent to which these algorithms can function in real world situations involving multiple stakeholders, such as the individuals at-risk or their clinicians, remains less understood. Given the sensitivities and high-stake decisions in the mental health domain, new human-centered approaches to the design, development, and evaluation of algorithms that detect mental illness risks are needed~\citep{chancellor2020methods,kim2021review}. 

Accordingly, we adopted guidance based on three core principles here: increasing agency of and provisioning power to the vulnerable; preventing unintended negative consequences of the use of these technologies; and conforming to ethical and compassionate approaches in the very design of the algorithms and models. To do so, we took inspiration from the action research framework~\citep{stringer2020action}, which is a philosophy and methodology in the qualitative social sciences. This approach seeks transformative change through the simultaneous process of taking action and doing research, which are linked together by critical reflection. 

\subsection{Personalized Forecasting of a Psychotic Relapse}
Toward greater stakeholder involvement in development of detection algorithms, in work involving intertwined collaboration with clinicians at Northwell Health, we have explored how online behavioral and activity based trace data, such as on social media, could provide a new ``lens'' and a form of collateral information in mental health treatment (see \citep{birnbaum2018digital,birnbaum2019detecting,birnbaum2020utilizing} as examples). Our long-term goal is to investigate if computational analyses of such data can  help us, by powering novel interventions and technologies, to reduce and even prevent significant morbidity and mortality, as well as ensure that at-risk individuals receive the timely care they need. In a formative study within this research, we sought to examine to what extent patient volunteered social media data can help clinical decision-making around relapse~\citep{birnbaum2019detecting}. Current strategies for relapse evaluation are severely limited by their reliance on direct, frequent, and timely contact with trained professionals as well as accurate and insightful patient and family recall~\citep{birnbaum2015role}. They do not allow for time-sensitive and objective monitoring of psychotic symptoms and lack the ability to detect subtle, sub-clinical, and burgeoning changes~\citep{barnett2018relapse,adler2020predicting}. 
Accordingly, in collaboration with clinical researchers and practitioners at Northwell Health, since 2018 we have recruited over 350 patients diagnosed with a psychotic disorder like schizophrenia, half of whom have had at least one relapse related hospitalization. All of these individuals agreed to share their entire Facebook archives with us, spanning 10+ years longitudinally, as well as their electronic medical records over 4+ years, opening up  opportunities for temporal and causal analyses of mental health at a scale and granularity rarely possible before. 

Adopting an iterative human-centered algorithm development process, we modeled the relapse prediction problem as an anomaly detection problem, where the idea was to identify person-centered deviations in behavior, mood, or activity that could be clinically meaningful from the perspective of the individual's mental health state and wellbeing. Using the relapse hospitalization dates as markers, we segmented each participant’s Facebook data into periods of relapse and periods of relative health. We trained a one-class classification Support Vector Machine algorithm on periods of relative health to identify distinguishing patterns of inliers. We then tested the best performing model on an unseen sample of both periods of relapse and relative health with the goal of predicting healthy periods as inliers and relapse periods as outliers. 
Our results were promising, indicating that our models could accurately detect a relapse 30 days in advance for 79\% of the cases (specificity of 71\% and a positive predictive value of 66\%). What these results tell us is that personalized methods to longitudinally forecast the likelihood of imminent adverse mental health outcomes, like a relapse, are feasible. 

This work, thus, has paved a way to provide the knowledge necessary to develop a new generation of innovative and targeted digital decision support tools that incorporate social media based algorithmic insights, eventually leading to large-scale trials exploring their acceptability and efficacy in clinical care. For instance, through this same collaboration, we have examined how the above inferred mental health status (e.g., likelihood of relapse), based on patients’ social media usage, can be useful collateral information which mental health clinicians and patients could utilize in a narrative- or evidence-based treatment paradigm~\citep{yoo2020designing}. Based on a speculative co-design approach, we have been designing novel interfaces that present these algorithmic inferences from patients’ social media data, to augment structured interviews and unstructured consultations by mental health clinicians~\citep{yoo2021clinician}. We have also identified how such indicators may enable and improve the design of patient-facing tools, such as those that promote self-reflection in management of mental illnesses~\citep{devakumar2021review}.

\subsection{Estimating Nationwide Suicide Rates}

In another work, which was a collaboration with the Centers for Disease Control and Prevention (CDC), we sought to inform suicide prevention efforts using social media in conjunction with other types of public health signals. As has now been discussed extensively in popular discourse, the rates of suicide in the US population have been increasing~\citep{curtin2016increase}. Given the urgency of this public health problem, there is a need for real-time information on suicide fatality trends to guide prevention efforts. However, such capability is currently lacking. With national statistics on suicide rates delayed by 1-2 years, currently it is particularly challenging for government agencies to make budgetary decisions, deploy interventions and adapt policies for improving timely access to care. To address this gap, we developed a first of its kind machine learning model to estimate weekly suicide fatalities in the U.S. in near real-time~\citep{choi2020development}. 

Specifically, we developed and validated a two-phase machine learning pipeline to leverage several disparate but complementary streams of real-time information related to suicide. These data streams encompassed novel health services administrative data, economic data, meteorological data, and online (social media, search engine) data. Basically first we focused on finding the best predictor of the weekly number of suicide fatalities for each given data source, by taking a training-validation approach. In particular, first, a model (for a given data source) was trained based on the time-series data from a single source (i.e., Twitter) and the weekly numbers of suicide fatalities. These individual models gave intermediate estimates (a prediction of the weekly number of suicide fatalities in the given week of interest for the same data source). Then the second phase involved combining these estimates via an ensemble approach (a deep neural network model) to arrive at a single estimate of weekly suicide fatalities that incorporated signals from all of the different data sources. 

Expectedly, we found all of the ensembles improve over models using individual data sources. Importantly, the ensemble outperformed predictions made from the best performing baseline model, which made estimates from historical suicide fatalities alone. 
Concretely, the ensemble model improved the correlation while also reducing error around weekly estimates by approximately half and the error for the annual estimate to less than one-tenth of that of the baseline model. 

There are two main takeaways from this research. First, as mentioned above, there is currently no established way to gather real-time national information on suicide trends, which is essential for timely suicide prevention efforts. The developed ensemble framework provides a mechanism to address this challenge. Second, our approach to use multiple, complementary datasets in the estimation task outperformed predictions made from the best performing baseline model, 
thus indicating that using multiple, complementary datasets is robust against ``big data hubris'' 
-- limitations noted in prior efforts~\citep{lazer2014parable} that have often undermined the practical utility of previous public health detection technologies. 

\section{Ethics}

As discussed above, algorithmic inferences derived from social media data of individuals hold great potential in supporting early detection and treatment of mental disorders and in the design of interventions. At the same time, the outcomes of this research can pose great risks to individuals, such as issues of incorrect, opaque algorithmic predictions, involvement of bad or unaccountable actors, and potential biases from intentional or inadvertent misuse of insights. Amplifying these tensions,
there are also divergent and sometimes inconsistent methodological gaps and under-explored ethics and privacy dimensions. In fact, through a systematic literature review~\citep{chancellor2020methods}, we found that most existing research in this area focused on abstract notions of ethics and methodological rigor to understand public health using social media data. As noted by other scholars as well, the use of social media for health research more broadly has been shown to raise a variety of ethical concerns surrounding users’ expectations regarding the distinction between public and private content~\citep{conway2016social}, privacy behaviors~\citep{fiesler2018we}, and researcher responsibilities~\citep{vitak2016beyond}. Specifically, while a significant body of this research relies on publicly accessible social media data, user perspectives need to be factored in, ranging from their views on data ownership to expectations of what is appropriate to be inferred; simply because data is public it does not indicate it is ethical to use such data to build prediction models~\citep{benton2017ethical,fiesler2018participant}. 

Motivated from this burgeoning research, we looked at how human data has been conceptualized in the literature that has employed social media to assess and predict mental and behavioral health outcomes~\citep{chancellor2019human}. We note that in machine learning and related areas, there is little guidance for managing researcher relationships to data; machine learning has historically relied on large and public benchmark datasets, like ImageNet~\citep{krizhevsky2012imagenet}. However, in a domain like mental health, data now comes from sources much closer to the human. 
As data-driven research moves closer to research traditionally protected through ethics boards~\citep{shilton2016we}, this new proximity complicates the traditional representations of
individuals involved in work from either machine learning or human subjects research perspectives alone. These representations of the human have downstream consequences on how research is conducted and reported, and how it may impact communities and individuals who are the object of study.

But the impacts of these representations go beyond abstract notions of roles or responsibilities. A systematic review of papers published on the topic of using social media to predict mental health indicated to us that computationally focused work tends to treat individuals as data points to be analyzed, rather than as a person who should be justly,
equitably, and sensitively treated, thus abstracting away from unique details to identify large-scale patterns and phenomena. Although predictions of mental health states can support decision-making and assist the design of intelligent interventions~\citep{de2014can}, as we argued elsewhere in this chapter, this systematic review underscored that such predictions can also provoke extensive utopian and dystopian rhetoric (as noted by other scholars as well~\citep{boyd2012critical,lazer2014parable}). Therefore, when the representations of data are dehumanizing, as we showed in this research, they can have serious repercussions for research methods and practical risks in increasing stigma, reproducing stereotypes and discriminatory practices, and harming individuals and communities.  In short, our findings surfaced a paradoxical representation of the human in this body of work, as being both in the ``subject'' and ``object'' positions, where humans are both centered and prioritized in the analyses but are also the object of machine learning techniques. 

Noting these challenges, we developed a taxonomy of issues in algorithmic prediction of mental health status on social media data.  First, we identified the gap between ethics committees and participants in such research~\citep{chancellor2019taxonomy}. 
Second, we identified tensions in methods, such as construct validity and bias, interpretability of algorithmic output, and privacy. Finally, we examined implications of this research in benefiting mental health research, challenges faced by key stakeholders, and the risks of designing interventions. In short, the taxonomy was used to derive several calls to action for the broader community: a) \textit{Participatory algorithm design}: Researchers should include key stakeholders in the research process, including clinicians, social networks, and individuals who are the object of these predictions; b) \textit{Developing best practices for methods}: In published work, researchers should disclose study design and methods decisions to promote reproducibility, and the field should agree on what best practices are; and c)  \textit{Beyond ethics boards}: Researchers need to consider and discuss the implications of this research, outside of the normal considerations of ethics committees, by incorporate ethics as a key value in the research process from the beginning. 

\section{Conclusion}

The ability to illustrate and model individual behavior using their social media data, that can predict their mental health state, shows promise in the design and deployment of next-generation wellness facilitating technologies~\citep{de2013role}. We envision privacy-preserving software applications and services that can serve as early warning systems providing personalized alerts and information to individuals~\citep{de2013moon}. Clinician-facing applications and dashboards may also serve unmet need by surfacing valuable collateral information that can aid in clinical decision-making~\citep{yoo2020designing}. And more, self-reflection and personal interfaces can help individuals with lived experience of mental illness to better manage their condition~\citep{taylor2021misfires}. Broadly speaking, these tools perhaps can enable adjuvant diagnosis and monitoring of mental illnesses, complementary to conventional approaches. 

Despite the contributions of the above research, several unanswered questions remain. A natural next question that arises from this research is whether the developed algorithms are ready for deployment, and what levels of algorithmic performance are required for real world translation. To answer this we would need to navigate the question of the gap that exists between academically acceptable gains in predictive performance and what a practical situation demands; scholars have critiqued that the practical gains of many machine learning algorithms do not necessarily improve much over the state-of-the-art~\citep{goel2010predicting}. For instance, a classifier that detects the relapse likelihood of schizophrenia with 85\% accuracy may not be helpful to a clinician who intends to use this algorithm to determine if, for a specific patient, there is a high likelihood of relapse in the near future. 
Complementarily, as we emphasize achieving greater algorithmic performance, it is important to acknowledge that machine learning models are inherently uncertain models and would never be perfect realistically. In that case, how do we support graceful failure when our data or models cannot stand up to the potential use case -- in other words, how can algorithms be designed to ``do no harm'', as the Hippocratic Oath goes~\citep{antoniou2010reflections}. 
Further, our recent research has shown that individual technology-based mental health interventions do not exist in a vacuum~\citep{pendse2020like}; 
rather they exist as part of a socially situated and structurally influenced pathway to care. For example, we found that people were dissuaded from calling mental health helplines after having poor experiences in therapy, and conversely, they were hesitant to try therapy after poor experiences on helplines. Therefore, as these algorithms make their way into technology-driven interventions, to support their real world translation, research is needed to identify ways to integrate the above detection and predictive algorithms into existing pathways that people pursue to find help and care, and how might the algorithms be useful to end users in the face of the barriers that exist in these pathways, whether subject to their identity or socio-economic-political context. 

Next there are questions about the future of mental health work. In a future where the ML algorithms are adopted to administer treatment to patients or support public health decision-making, how can we ensure that they are compatible with the existing work practices of the human experts? Ensuring this compatibility is important to guard against challenges of algorithmic or AI mistrust as well as over-reliance, for supporting a balance between agency of the experts and automation of aspects of the decision-making, and to facilitate building shared mental models between the offerings and strengths of the human expert and the algorithm~\citep{lee2021artificial}. Integrating these algorithms in the existing work practices, however, begs the question of skill acquisition -- would we require the human experts to now also understand how ML/AI models work and how they should decipher the algorithms? What type of additional burden would the intent to use these algorithms impose on these human experts? Importantly, as humans and algorithms share initiatives in such a future of work with a mixed initiative technology design~\citep{horvitz1999principles}, legal and ethical considerations surrounding the ``duty to rescue'' would also have to be navigated. Finally, introducing algorithms within the decision-making process is likely to impact human-human relationships, as our recent research has revealed~\citep{pendse2021can}, whether patient-clinician therapeutic alliance or coordination among different public health actors. Therefore, as we think about real world deployed ML/AI systems that would function over time, we have to recognize that sustained social relationships are important, including protecting them against strained interactions and compromised interpersonal boundaries. 

In conclusion, we have to recognize we are not just aiming to develop the most accurate machine learning algorithm or envisioning how AI and data may transform the status quo in mental health, but acknowledge that we are tackling a socio-technical problem. It is a socio-technical problem because of its social piece -- how can we humanize our approaches so that those struggling with mental illness, or those vulnerable and marginalized, have an opportunity to work with researchers and stakeholders to define what may work best for them? What would it mean it to incorporate their voices, needs, desires, and demands in the very design and development of the algorithms?

\section{Further Readings}

See \citet{chancellor2020methods} and \citet{guntuku2017detecting} for systematic reviews on this topic; \cite{eichstaedt2018facebook} who presented a first study that linked Facebook derived signals of depression with diagnoses noted in patients' electronic medical records; and \cite{benton2017ethical} and \cite{chancellor2019taxonomy} for ethical reflections.

\clearpage 
\bibliographystyle{plainnat}

\end{document}